\documentstyle[12pt]{article}

\begin{document}

\pagestyle{empty}

\hfill {\bf McGill--96/38}

\hfill {\bf UA/NPPS--7--96}

\hfill {\bf hep-ph/9611401}

\hfill September 1996

\vglue 2cm

\begin{center} \begin{Large} \begin{bf}
Direct photon production in polarized hadron collisions at collider 
and fixed target energies
\end{bf} \end{Large} \end{center}
\vglue 0.35cm
{\begin{center}
A.P.\ Contogouris$^{a,b,1}$
and Z.\ Merebashvili$^{a,*,2}$ \end{center}}
\parbox{6.4in}{\leftskip=1.0pc
{\it a.\ Department of Physics, McGill University, Montreal,
Qc., H3A 2T8, Canada}\\
\vglue -0.25cm
{\it b.\ Nuclear and Particle Physics, University of Athens,
Athens 15771, Greece}
}
\begin{center}
\vglue 1.0cm
\begin{bf} ABSTRACT \end{bf}
\end{center}
%\vglue 1.0cm
{%\rightskip=1.5pc
 %\leftskip=1.5pc
 %\tenrm\baselineskip=12pt
 \noindent
Recently determined next-to-leading order sets of polarized parton
distributions are used to study large-$p_T$ $\vec{p}\vec{p}\rightarrow
\gamma+X$ at $\sqrt{s}=38$, 100 and 500 GeV. Certain conversion
terms, necessary to use the above sets, are determined. It is concluded 
that, to distinguish between the above sets, planned RHIC experiments 
should be successful, in particular at c.m. photon pseudorapidity
$\simeq 1$, and that a proposed HERA--$\vec{N}$ fixed target
experiment should have rather large accuracy at relatively large
$x_T \; (=2p_T/\sqrt{s})$.
}

\renewcommand{\thefootnote}{*}
\footnotetext{On leave from High Energy Physics Institute, Tbilisi
State University, Tbilisi, Republic of Georgia.}
\renewcommand{\thefootnote}{\arabic{footnote}}
\addtocounter{footnote}{1}
\footnotetext{e-mail: apcont@hep.physics.mcgill.ca}
\addtocounter{footnote}{1}
\footnotetext{e-mail: zaza@hep.physics.mcgill.ca}

\newpage

\pagestyle{plain}
\setcounter{page}{1}

%\vglue .3cm
\begin{center}\begin{large}\begin{bf}
I. INTRODUCTION
\end{bf}\end{large}\end{center}
\vglue .3cm

Polarized particle reactions have opened a new domain to test  
perturbative
QCD. In this domain, in spite of extensive recent data on polarized deep 
inelastic scattering $[$\ref{r1}$]$, the shape and the size of the  
polarized
gluon distribution $\Delta g(x)$ remains an open question. Important 
progress requires experiments on reactions with polarized initial  
hadrons,
dominated by subprocesses with initial gluons, like those planned in 
$[$\ref{r2}$]$ or proposed in $[$\ref{r3}$]$.

Among such reactions, large-$p_{T}$ direct photon production in
longitudinally polarized proton-proton collisions occupies a prominent 
place, hence higher (next-to-leading) order corrections (HOC) have been 
determined $[$\ref{r4},\ref{r5}$]$. The results, based on the then  
existing
leading order polarized parton distributions, predicted large  
asymmetries.

Recently, however, there have been important developments in two  
respects.
First, the two-loop longitudinally polarized splitting functions became 
known $[$\ref{r6}$]$. Second, making use of them as well as of the  
recent data
$[$\ref{r1}$]$, sets of new polarized parton distributions have  
been determined.
As before, the sets differ essentially in the input $\Delta g(x)$. In 
contrast, however, to older sets, due to $[$\ref{r1}$]$, the new sets 
exclude very large values of the integral $\Delta G \equiv \int_{0}^{1} 
\Delta g(x)\,dx$. Thus the predictions of $[$\ref{r4},\ref{r5}$]$  
must be
reconsidered, and this is the main purpose of the present work.

Regarding the extension of the Dirac matrix $\gamma_{5}$ in $n=4-2
\varepsilon$ dimensions, in our work $[$\ref{r4}$]$ we adopted an
anticommuting $\gamma_{5}$ (AC) scheme as well as dimensional reduction 
(see below). The two-loop splitting functions of $[$\ref{r6}$]$ are 
determined in a different scheme, and one must calculate the terms to 
be added to the hard scattering cross section of $[$\ref{r4}$]$ to  
convert
to the scheme of $[$\ref{r6}$]$ (conversion terms). An additional  
purpose
of this work is to determine these terms.

In Sect.~2 we determine the conversion terms. In Sect.~3, using three 
new sets of polarized parton distributions, we present detailed results 
and compare them to $[$\ref{r4}$]$. Sect.~4 contains our conclusions.

\renewcommand{\theequation}{2.\arabic{equation}}
\vglue 1cm
\begin{center}\begin{large}\begin{bf}
II. CONVERSION TERMS
\end{bf}\end{large}\end{center}
\vglue .3cm

In a given scheme, the $n$-dimensional split function $P_{ba}^{n}(z,
\varepsilon)$ is written
\begin{equation}
P_{ba}^{n}(z,\varepsilon)=P_{ba}(z)+\varepsilon P_{ba}^{\varepsilon}(z);
\end{equation}
the function $P_{ba}^{\varepsilon}(z)$ will be termed  
$\varepsilon$-part.
In the same scheme, at some factorization scale $M$, let $f_{a/A}(x,M)$ 
denote parton distribution. In a different scheme, the corresponding 
quantities are distinguished by primes. Then the parton distributions 
transform according to: $[$\ref{r7}$]$
\begin{equation}
f_{a/A}^{\prime}(x,M) = f_{a/A}(x,M) + \frac{\alpha_{s}(\mu)}{2\pi}
\int_{x}^{1} \frac{dy}{y} f_{a/A}(y,M) \left[P_{ba}^{\prime \varepsilon}
\left(\frac{x}{y}\right) - P_{ba}^{\varepsilon}\left(\frac{x}{y}\right)
\right] + {\cal O}(\alpha_{s}^{2}),
\end{equation}
where $\mu$ is a renormalization scale. Polarized parton distributions, 
$\Delta f_{a/A}(x,M)$, transform in a similar manner in terms of the 
polarized $\varepsilon$-parts, $\Delta P_{ba}^{\varepsilon}(z)$.

In view of the multitude of schemes, in particular for problems of
polarized particles, we specify those relevant to the present work.

The first is the anticommuting $\gamma_5$ scheme previously
mentioned. This was introduced in $[$\ref{r8}$]$ and was used to
determine the HOC of $[$\ref{r8}$]$ and most of the HOC of  
$[$\ref{r4}$]$.
In this scheme, in addition to $\Delta P_{qq}^{n}(z,\varepsilon) \left
(=P_{qq}^{n}(z,\varepsilon)\right)$, we need and have determined
$\Delta P_{gg}^{n}(z,\varepsilon)$ and $\Delta P_{qg}^{n}(z,
\varepsilon)$. The complete expressions are:
\begin{eqnarray}
\label{e2.3}
\nonumber
\Delta P_{qq}^{n}(z,\varepsilon)&=&C_F\left\{\frac{1+z^2}{(1-z)_+}
-\varepsilon (1-z) + \frac{3+\varepsilon}{2} \delta(1-z)
\right\}  \\  \nonumber
\Delta P_{gg}^{n}(z,\varepsilon)&=&2N_C\left\{\frac{1}{(1-z)_+}-2z+1
\right\}+\left(b+\varepsilon\frac{N_F}{6}\right) \delta(1-z)  \\
\Delta P_{qg}^{n}(z,\varepsilon)&=&2N_F(z-\frac{1}{2})
\mbox {\hspace{1.2in} (AC scheme)}
\end{eqnarray}
where $C_F=4/3$ and $N_C=3$ (colour SU(3)), $N_F$ is the number of
flavours and $b=(11N_C-2N_F)/6$.

The second scheme is dimensional reduction (RD) $[$\ref{r9}$]$, in which 
part of the HOC of $[$\ref{r4}$]$ were determined. In RD, for all a,b:
\begin{equation}
\label{e2.4}
P_{ab}^{\varepsilon}(z)=\Delta P_{ab}^{\varepsilon}(z)=0
\mbox {\hspace{1.2in} (RD scheme)}
\end{equation}

The third is the t'Hooft--Veltman (HV) scheme $[$\ref{r10}$]$,  
apart from
the $\varepsilon$-part of $\Delta P_{qq}^n$, which is taken as in
(\ref{e2.3}) to satisfy helicity conservation. Here we need all split 
functions, and the remaining ones are found to be (see also the
second of $[$\ref{r6}$]$):
\begin{eqnarray}
\label{e2.5}
\nonumber
\Delta P_{gg}^{n}(z,\varepsilon)&=&2N_C\left\{\frac{1}{(1-z)_+}-2z+1
+2\varepsilon(1-z)
\right\}+\left(b+\varepsilon\frac{N_F}{6}\right) \delta(1-z)  \\
\Delta P_{qg}^{n}(z,\varepsilon)&=&2N_F\left\{z-\frac{1}{2}-\varepsilon
(1-z)\right\}  \nonumber  \\
\Delta P_{gq}^{n}(z,\varepsilon)&=&C_F\{2-z+2\varepsilon(1-z)\}
\mbox {\hspace{1.2in} (HV scheme)}
\end{eqnarray}
It is this scheme to which we should convert our results.

The conversion terms are known to arise from differences in the
$\varepsilon$-parts of the split functions and are completely determined 
from the factorization counterterms $[$\ref{r7}$]$. The form of the  
latter
has been given in $[$\ref{r8}$]$, and the determination of the  
conversion
terms is straightforward.

Denoting a given subprocess by $a(p_1)+b(p_2)\rightarrow\gamma(p)+
c+d$, where the quantities in parentheses are 4-momenta, we define:
\[  \hat{s}=(p_1+p_2)^2 {\rm \hspace{.4in}}  \hat{t}=(p_1-p)^2
{\rm \hspace{.4in}}  \hat{u}=(p_2-p)^2    \]
and
\begin{equation}
\label{e2.6}
v=1+\hat{t}/\hat{s} {\rm \hspace{.8in}} w=-\hat{u}/(\hat{s}+\hat{t})
\end{equation}
Below we give the terms to be added to $\Delta f=\Delta f(v,w)$ defined 
by: $\{$Eq. (4.4) of $[$4(b)$]\}$
\begin{eqnarray}
\label{e2.7}
\nonumber
E\frac{\Delta d\sigma}{d^{3}p}&=&\frac{\Phi}{\pi p_{T}^{4}}
\int_{v_1}^{v_2}\!\!dv\,v(1-v)\int_{w_1}^{1}\!\!dw\,w\Delta F_{a/A}(x_a)
\Delta F_{b/B}(x_b)\{\Delta B(v)\delta (1-w)+\frac{\alpha_s}{2\pi}
\Delta f\}   \\
& & {\rm \hspace{3in}} +(A\leftrightarrow B, \; \eta\leftrightarrow  
-\eta)
\end{eqnarray}
where $v_1,v_2,w_1,x_a=x_a(v,w), x_b=x_b(v,w)$ and $\Delta B$ are given 
in $[$4(b)$]$; $\eta$ is the c.m. pseudorapidity of the photon and
$\Phi$ is also specified below.

First consider $\vec{g}(p_1)+\vec{q}(p_2)\rightarrow\gamma(p)+q+g$. 
We need the conversion term from AC to HV scheme. With $\Phi=\pi
\alpha\alpha_s e_{q}^{2}/N_C$, writing
\[ A\equiv v(1-v)w/2, \]
and setting $2N_F\rightarrow1$ in the expressions of
$\Delta P_{qg}^{n}(z,\varepsilon)$ we find:
\begin{equation}
\label{e2.8}
\Delta f_{qg}=-A^{-1}(1-w)\{C_F[v^2+(1-v)^2]+2N_C(1+v)(1-v)^2\}
\end{equation}

For $\vec{g}\vec{g}\rightarrow\gamma q\bar{q}$ we should also go from 
AC to HV scheme. With $\Phi=\pi\alpha\alpha_se_{q}^{2}/8$:
\begin{equation}
\label{e2.9}
\Delta f_{gg}=2C_FA^{-1}v^2(1-w)(3-v+vw)
\end{equation}

Now consider $\vec{q}_{\alpha}(p_1)+\vec{q}_{\beta}(p_2)\rightarrow
\gamma(p)+q_{\alpha}+q_{\beta}$. We need the conversion term from
RD to HV scheme. Here the $\varepsilon$-part of the split function
$P_{\gamma q}^n$ in the (conventional) dimensional regularization
scheme is required. We find:
\[  P_{\gamma q}^{n}(z,\varepsilon)=[1+(1-z)^2]/z-\varepsilon z
\mbox {\hspace{.4in}     (dim. reg.)}    \]
Then, with $\Phi=\pi C_F\alpha\alpha_s/2N_C$, we find for
$\beta\neq\alpha$:
\begin{eqnarray}
\label{e2.10a}
\nonumber
\Delta f_{qq}=2v\{e_{\alpha}^2[1+\frac{2vw}{1-v}-
\frac{1-w}{A}\frac{v^3w^2(2-vw)}{(1-vw)^2}]
+e_{\beta}^2[1+\frac{2(1-v)}{vw}    \\
{\rm \hspace{2in}} -\frac{1-w}{vA}(1+v)(1-v)^2]\}
\end{eqnarray}
and for $\beta=\alpha$ and $z=1-v+vw$:
\begin{eqnarray}
\label{e2.10b}
\nonumber
\Delta f_{qq}=2ve_{\alpha}^2\{2[1+\frac{vw}{1-v}+\frac{1-v}
{vw}-\frac{1}{N_C}\frac{z^2}{vw(1-v)}]-\frac{1-w}{A}[
\frac{v^3w^2(2-vw)}{(1-vw)^2}   \\
{\rm \hspace{2in}}  +\frac{(1+v)(1-v)^2}{v}]\}
\end{eqnarray}

Finally we consider $\vec{q}(p_1)+\vec{\bar{q}}(p_2)\rightarrow\gamma 
(p)+q+\bar{q}$ ($q,\bar{q}$ of identical flavour) and give the
conversion term from RD $\rightarrow$ HV scheme. With $\Phi=\pi
C_F\alpha\alpha_s/2N_C$:
\begin{eqnarray}
\nonumber
\label{e2.11}
\Delta f_{q\bar{q}}=2ve_q^2\{1+\frac{2vw}{1-v}-\frac
{(1-v)^2+v^2w^2}
{z^2}-\frac{2}{N_C}\frac{v^2w^2}{(1-v)z}-\frac{1-w}{A}[
\frac{v^3w^2(2-vw)}{(1-vw)^2}  \\
{\rm \hspace{2in}}   +\frac{(1+v)(1-v)^2}{v}]\}
\end{eqnarray}

The contributions of $q\bar{q}\rightarrow\gamma gg$ and $\vec{q}_
{\alpha}\vec{\bar{q}}_{\alpha}\rightarrow\gamma q_{\beta}\bar{q}_{\beta}
$ with $q_{\beta}\bar{q}_{\beta}$ produced via $g\rightarrow
q_{\beta}\bar{q}_{\beta}$ are obtained from the corresponding
unpolarized subprocesses by a change of sign. As in $[$\ref{r4}$]$, we 
use unpolarized results in dimensional regularization  
($\overline{\rm \rm MS}$
scheme) and in view of taking $\Delta P_{qq}^n=P_{qq}^n$ (see before 
Eq.~(\ref{e2.5})), conversion terms are absent.

Note that (\ref{e2.9}) is the same as the corresponding term of
$[$\ref{r11}$]$, a fact easily understood: The terms of  
$[$\ref{r11}$]$ convert
from the so-called $\overline{\rm MS}_P$ scheme, introduced in  
$[$\ref{r5}$]$,
to HV. The $\overline{\rm MS}_P$ scheme has $\Delta P_{qq}^{\varepsilon}=
C_F[-(1-z)+\delta(1-z)/2], \Delta P_{gg}^{\varepsilon}=N_F\delta (1-z)
/6$ and $\Delta P_{ab}^{\varepsilon}=0$ for the rest. Thus, as far as 
we are concerned, it is equivalent to the AC scheme. Likewise, in
(\ref{e2.10a})--(\ref{e2.11}), the parts $\sim 1/A$ are the same as 
the corresponding terms of $[$\ref{r11}$]$; they amount to converting 
from $\overline{\rm MS}_P$ to HV. Finally, with the transformation
$v^{\prime}=1-vw, w^{\prime}=(1-v)/(1-vw) \; [{\rm i.e.}\;\hat{t}
\leftrightarrow\hat{u}]$, and taking into account that the Jacobian 
$\partial (v^{\prime},w^{\prime})/\partial (v,w)=v/(1-vw)$, we find that 
(\ref{e2.9}) becomes the same as the corresponding result of
$[$\ref{r11}$]$; it should be so, since the latter interchanges  
$p_1$ and
$p_2$ between initial partons.

Some doubts have been expressed about the possibility to convert from 
AC to HV scheme $[$\ref{r11}$]$. In view of the above, and as far as we 
can tell, the doubts are unfounded.

\renewcommand{\theequation}{3.\arabic{equation}}
\setcounter{equation}{0}
\vglue 1cm
\begin{center}\begin{large}\begin{bf}
III. RESULTS AND DISCUSSION
\end{bf}\end{large}\end{center}
\vglue .3cm

We use various sets of polarized parton distributions of {\em one}
group (the next-to-leading order sets A, B, C of $[$\ref{r12}$]$);
different groups proceed with different input assumptions, and
this may obscure the degree of real difference in the gluon
distribution.

Furthermore, in presenting asymmetries we always divide the
polarized by unpolarized cross sections determined via {\em one}
set of unpolarized parton distributions, of STEQ4M $[$\ref{r13}$]$; 
dividing by cross sections determined via distributions of
different sets obscures to some extend the differences in
$\Delta g(x)$. CTEQ4M has $\Lambda_4=0.296$ GeV, which we also use
in our (two-loop) expression of $\alpha_s(\mu)$, varying it as
we cross the $b$-quark threshold ($m_b=4.5$ GeV).

An explanation of our attitude towards $\gamma$ Brems and the related 
$\gamma$ fragmentation is in order. First, the related
effects are known to be important at rather small $x_T(=2p_T/\sqrt{s})$. 
The RHIC experiment, being a colliding beam one, is expected to
observe {\em isolated} $\gamma$'s. Then, unless isolation criteria
are specified, a calculation of $\gamma$ Brems is of dubious value. 
The proposed HERA--$\vec{N}$ experiment $[$\ref{r3}$]$, being a
fixed--target one, may observe non-isolated $\gamma$'s. However,
it's energy ($\sqrt{s}=39$ GeV) is relatively low and the values of 
$x_T$ for which $\gamma$ Brems is important correspond to $p_T$
rather small. At small $p_T$, other effects, like intrinsic parton's 
$k_T$, higher twist etc also become important. In particular at
lower $\sqrt{s}$, the cross sections are steeper and $k_T$ effects
are stronger.

We thus prefer to leave out $\gamma$ Brems; as in $[$\ref{r4}$]$, our 
calculations include only the factorization counterterms necessary
to cancel the mass singularities of collinear $\gamma$ emission. The 
subsequent results correspond to $x_T>0.08$.

We briefly comment on the input $\Delta g(x)$ of $[$\ref{r12}$]$  
($Q_0^2=4
{\rm GeV}^2$). In sets A, B, $\Delta g(x)>0$ throughout; in set C,
$\Delta g(x)$ changes sign, and for $x>0.1$ becomes negative. The
integral $\Delta G$ has its largest value for A and its smallest
for C; even for A, however, it is significantly smaller than the
large $\Delta G$ of $[$\ref{r14}$]$ used in $[$\ref{r4}$]$.

As in $[$\ref{r4}$]$, we present results for $\vec{p}\vec{p}\rightarrow 
\gamma+X$ at $\sqrt{s}=38$, 100 and 500 GeV; the first value is
relevant to the HERA--$\vec{N}$ experiment $[$\ref{r3}$]$. Also, we 
consider photon c.m. pseudorapidities $\eta=0$, 1 and 1.6 and use
$\mu=M=p_T$. As there are similarities in the pattern, in Figs.~1
and 2(a), (b), we present results for set A and $\eta=1$, but we
comment on the results for other sets and $\eta$'s.

Beginning with $\vec{q}\vec{g}\rightarrow\gamma q$, denote by
$\sigma_{B}(gq)\,[\sigma_{HO}(gq)]$ the contribution of $\Delta B\,
[\Delta f]$ to $E\Delta d\sigma/d^3p$, Eq.~(\ref{e2.7}). Compared
to $[$\ref{r4}$]$, in the considered range of $x_T$, for sets A, B the 
$K$--factor $K_{gq}\equiv[\sigma_{B}(gq)+\sigma_{HO}(gq)]/\sigma_{B}
(gq)$, changes very little; this is clear in Fig.~1(a). For set C,
at $\sqrt{s}=38$ and 100 GeV and $\eta=1.6$ and 1, $K_{gq}$ is not
smooth; at $\eta=0$ and for $\sqrt{s}=500$ at all $\eta$, it is
similar to Fig.~1(a).

Next, for $\vec{g}\vec{g}\rightarrow\gamma q\bar{q}$ consider the
corresponding $\Delta f$ and denote by $\sigma(gg)$ its contribution 
to $E\Delta d\sigma/d^3p$. The ratio $K_{gg}\equiv\sigma(gg)/\sigma
_{B}(qg)$, used as a measure, for sets A, B is negative and in
magnitude very small over all our kinematic range; Fig.~1(b) presents 
a typical $K_{gg}$. For C it is similar, except for $\sqrt{s}=38$
and 100 GeV and small $x_T$, where $K_{gg}>0$, but still very small.

For $\vec{q}\vec{q}\rightarrow\gamma qq$, denoting by $\sigma(qq)$
the contribution of the corresponding $\Delta f$, we consider
$K_{qq}\equiv\sigma(qq)/\sigma_{B}(qg)$. For all sets, $K_{qq}$ is
negative. In magnitude, for sets A, B it is very small, for C somewhat 
larger. Fig.~1(c) shows a typical $K_{qq}$.

Now we consider the $K$-factor for all ${\cal O}(\alpha_s)$ and
${\cal O}(\alpha_s^2)$ contributions:
\begin{equation}
\label{e3.1}
K\equiv[\sigma_{B}(gq)+\sigma_{HO}(gq)+\sigma_{B}(q\bar{q})+\sigma_{
HO}(q\bar{q})+\sigma(gg)+\sigma(qq)]/[\sigma_{B}(gq)+\sigma_{B}(q
\bar{q})]
\end{equation}
For sets A, B, due to smallness of $|K_{gg}|$ and
$|K_{qq}|$, $K$ is almost the same as $K_{qg}$; Fig.~2(a) makes
this clear. As for $K_{qg}$, for set C, $K$ is smooth only at $\sqrt
{s}=500$ GeV.

Turning to cross section $E\Delta d\sigma/d^3p$, for sets A, B the
shapes are similar to $[$\ref{r4}$]$ but the magnitudes smaller; this is 
clear in Figs. 2(b) and 2(c). We remark $[$Fig.~2(c)$]$ that, as in 
$[$\ref{r4}$]$, the maximum of $E\Delta d\sigma/d^3p$ is at $\eta>0$, 
near $\eta=1$ at the higher energies. For set C the behaviour is
complicated: at $\sqrt{s}=38$ GeV for $\eta=0,1$ and 1.6, at $\sqrt
{s}=100$ for $\eta=0$ and 1 and at $\sqrt{s}=500$ for $\eta=0$,
$E\Delta d\sigma/d^3p$ changes sign as $x_T$ varies.

Figs.~3,4 present in detail asymmetries
\begin{equation}
\label{e3.2}
A_{LL}(p_{T},s,\eta)=E\Delta d\sigma(p_{T},s,\eta)/d^3p
\mbox{\large /} Ed\sigma(p_{T},s,\eta)/d^3p
\end{equation}
Considering first $\sqrt{s}=38$ GeV, for $p_T\leq6$ GeV, due to the 
smallness of $\Delta G$, $A_{LL}$ is small for all sets. It becomes 
larger at larger $p_T$, and for $x_T\geq0.4$ there is a clear
distinction, at least between set A and sets B, C\@. The cross
sections, however, become smaller and the proposed HERA--$\vec{N}$
experiment $[$\ref{r3}$]$ will need high accuracy. At $\sqrt{s}=100$ 
and 500 GeV (RHIC Spin), to distinguish between A and B, C should not 
be difficult at either $\eta\simeq 0$ or $\eta\simeq 1$; to  
distinguish between
B and C, data near $\eta\simeq 1$ (Fig.~4) may be necessary.

Note that, in general, $A_{LL}$ determined with HOC and without (only 
Born) differ. Fig.~4 shows the latter for set A, $\sqrt{s}=100$
and $\eta=1$ (dash-dotted line); at the higher $p_T$, it differs by 
a factor $\sim 2$.

Finally we turn to the effect of changing the scales, and consider
the ratio (Fig.~5, set A):
\begin{equation}
\label{e3.3}
r=E\Delta d\sigma(M=\mu=p_T/2)/d^3p
\mbox{\large /} E\Delta d\sigma(M=\mu=2p_T)/d^3p
\end{equation}
As one expects, with the evolution of the distributions determined
via two-loop splitting functions, there is more stability than
in $[$\ref{r4}$]$. With set B, $r$ is even more stable. Set C leads to 
unstable $r$, ranging from negative to positive values.

One may wonder about the effect of the conversion terms (Sect.~2) on 
our results. For all quantities considered, the effect does not
exceed 6\%; of course, mathematical consistency requires the presence 
of the terms.

\vglue 1cm
\begin{center}\begin{large}\begin{bf}
IV. CONCLUSIONS
\end{bf}\end{large}\end{center}
\vglue .3cm

Our essential conclusions, based on the next-to-leading order sets  
A, B, C of
$[$\ref{r12}$]$, are as follows:

In general, there is a marked difference between the predictions of  
sets A, B
$(\Delta g(x)>0)$ and of set C ($\Delta g(x)$ changing sign).

For sets A, B, the $K$-factors (\ref{e3.1}) exceed unity; thus the  
HOC enhance
the Born cross sections. For set C, $K$-factors are more  
complicated and in
several cases the HOC are of opposite sign to the Born.

Due to smaller $\Delta G$, the cross sections and asymmetries  
$A_{LL}$ are
smaller than in $[$\ref{r4}$]$. $A_{LL}$ become significant at  
relatively large
$p_T$, and the differences between $A_{LL}$ of the above three sets are 
larger at $\eta\simeq 1$, where also polarized cross sections are  
larger, in
general. Thus, to distinguish between A, B and C, RHIC, with its  
expected
high luminosity, should be successful, in particular near
$\eta=1$; on the other hand, HERA--$\vec{N}$ will need rather
high statistics at $x_T\geq 0.4$.

\vglue 1cm
\begin{center}\begin{large}\begin{bf}
ACKNOWLEDGEMENTS
\end{bf}\end{large}\end{center}
\vglue .3cm
We thank B.~Kamal for collaboration in determining the  
$\varepsilon$-parts of the
split functions of Sect.~2 and for several discussions, and  
K.~Fisher for drawing
the figures. This work was also supported by the
Natural Sciences and Engineering Research Council of Canada and
by the Quebec Department of Education.

%\newpage
\vglue 1cm
\begin{center}\begin{large}\begin{bf}
REFERENCES
\end{bf}\end{large}\end{center}
\vglue .3cm

   \begin{list}{$[$\arabic{enumi}$]$}
    {\usecounter{enumi} \setlength{\parsep}{0pt}
     \setlength{\itemsep}{3pt} \settowidth{\labelwidth}{(99)}
     \sloppy}
\item \label{r1}
SM Collaboration: B.~Adeva et al, Phys. Lett. {\bf B302}, 533 (1993); 
{\bf B320}, 400 (1994); D.~Adams et al, {\em ibid} {\bf B329},
399 (1994); {\bf B339}, 332 (1994) (E); {\bf B357}, 248 (1995).
\ \ SLAC--143 Collab.: D.~Anthony et al, Phys. Rev. Lett. {\bf 71}, 
959 (1993); K.~Abe et al, {\em ibid} {\bf 74}, 346 (1995);
{\bf 75}, 25 (1995); {\bf 76}, 587 (1996).
\item \label{r2}
RHIC, Phenix and Star Spin Collaborations ''Proposal on Spin Physics 
Using the RHIC Polarized Collider'' (Update), 1993.
\item \label{r3}
W.--D.~Nowak, preprint DESY 96--095.
\item \label{r4}
A.P.~Contogouris, B.~Kamal, Z.~Merebashvili and F.V.~Tkachov, (a)
Phys. Lett. {\bf B304}, 329 (1993); (b) Phys. Rev. {\bf D48},
4092 (1993); Erratum (in press).
\item \label{r5}
L.~Gordon and W.~Vogelsang, Phys. Rev. {\bf D48}, 3136 (1993);
{\bf D49}, 170 (1994).
\item \label{r6}
R.~Mertig and W.~van~Neerven, Z. Phys. {\bf C70}, 637 (1996);
W.~Vogelsang, preprint RAL-TR-96-020.
\item \label{r7}
B.~Kamal, Phys. Rev. {\bf D53}, 1142 (1996). See also J.~K\"{o}rner 
and M.~Tung, Z. Phys. {\bf C64}, 255 (1994).
\item \label{r8}
A.P.~Contogouris, S.~Papadopoulos and F.V.~Tkachov, Phys. Rev.
{\bf D46}, 2846 (1992).
\item \label{r9}
W.~Siegel, Phys. Lett. {\bf B84}, 193 (1979); I.~Jack et al, Z. Phys. 
{\bf C62}, 161 (1994); {\bf C63}, 151 (1994); M.~Ciuchini et al,
Nucl. Phys. {\bf B421}, 41 (1994).
\item \label{r10}
G.'t~Hooft and M.~Veltman, Nucl. Phys. {\bf B44}, 189 (1972).
\item \label{r11}
L.~Gordon and W.~Vogelsang, preprint ANL--HEP--PR--96--15.
\item \label{r12}
T.~Gehrmann and W.J.~Stirling, Phys. Rev. {\bf D53}, 6100 (1996).
\item \label{r13}
H.~Lai et al, preprint CTEQ--604.
\item \label{r14}
G.~Altarelli and W.J.~Stirling, Particle World {\bf 1}, 40 (1989).

\end{list}

%\newpage
\vglue 1cm
\begin{center}\begin{large}\begin{bf}
FIGURE CAPTIONS
\end{bf}\end{large}\end{center}
\vglue .3cm

\begin{list}{Fig.~\arabic{enumi}.}
    {\usecounter{enumi} \setlength{\parsep}{0pt}
     \setlength{\itemsep}{3pt} \settowidth{\labelwidth}{Fig.~9.}
     \sloppy
    }
\item{} Relative contributions of the basic subprocesses using the
distributions of set A at $\eta=1$. Dashed lines: $\sqrt{s}=38$ GeV, 
dotted: $\sqrt{s}=100$, solid: $\sqrt{s}=500$. (a) The $K$-factor  
$K_{gq}$
for $\vec{g}\vec{q}\rightarrow\gamma q$.\ (b) The factor $K_{gg}
\equiv\sigma(gg)/\sigma_B(gq)$ for $\vec{g}\vec{g}\rightarrow\gamma 
q\bar{q}$.\ (c) The factor $K_{qq}\equiv\sigma(qq)/\sigma_B(gq)$ for 
$\vec{q}\vec{q}\rightarrow\gamma qq$.
\item{} Results with set A: (a) The $K$-factor (\ref{e3.1}) for all  
${\cal O}
(\alpha_s)$ and ${\cal O}(\alpha_s^2)$ contributions at $\eta=1$.\ (b) 
Inclusive cross sections for $\vec{p}\vec{p}\rightarrow\gamma+X$ vs 
$x_T=2p_T/\sqrt{s}$ for $\eta=1$. Lines as in Fig.~2(a).\ (c) Inclusive 
cross sections for $\vec{p}\vec{p}\rightarrow\gamma+X$ vs $\eta$.
\item{} Asymmetries $A_{LL}$ for all three sets of distributions at 
pseudorapidity $\eta=0$.
\item{} As Fig.~3 for $\eta=1$.
\item{} Ratio $r\equiv[E\Delta d\sigma(M=\mu=p_T/2)/d^3p]
\mbox{\large /} [E\Delta d\sigma(M=\mu=2p_T)/d^3p]$ with set A.
Solid lines: $\eta=1.6$. Dotted: $\eta=1$. Dashed: $\eta=0$.
\end{list}

\end{document}